\documentclass[ejsv2]{imsart}

\RequirePackage[authoryear]{natbib}
\RequirePackage[colorlinks,citecolor=blue,urlcolor=blue]{hyperref}
\RequirePackage{graphicx}

\usepackage[english]{babel}
\usepackage{amscd, bm, bbm}   
\usepackage{enumerate}
\usepackage[shortlabels]{enumitem}

\theoremstyle{plain}
\newtheorem{thm}{Theorem}
\newtheorem{cor}{Corollary}

\newtheorem{prop}{Proposition}

\newtheorem{rem}{Remark}

\newcommand{\ip}[1]{\left\langle #1 \right\rangle} 

\newcommand{\argmax}{\operatornamewithlimits{argmax}} 

\newcommand{\tr}{\mathrm{tr}} 

\newcommand{\cond}{\xrightarrow{\text{d}}} 

\newcommand{\eo}{\mathsf{E}}

\newcommand{\Dt}{\Delta} 
\newcommand{\e}{\varepsilon} 
\newcommand{\ld}{\lambda} 

\newcommand{\I}{\mathbb{I}} 
\newcommand{\R}{\mathbb{R}} 

\newcommand{\mbs}{\mathbf{s}} 
\newcommand{\mby}{\mathbf{y}} 
\newcommand{\mbtheta}{\mathbf{\bm{\theta}}}

\newcommand{\kll}{\mathcal{L}}	
\newcommand{\ttt}{\mathcal{T}}	

\newcommand*{\normal}{\mathsf{N}}

\makeatletter
\newcommand{\leqnomode}{\tagsleft@true}
\newcommand{\reqnomode}{\tagsleft@false}
\makeatother

\arxiv{2010.00000}
\startlocaldefs
\theoremstyle{plain}

\theoremstyle{remark}

\endlocaldefs

\begin{document}
\begin{frontmatter}
\title{Generalized linear models with spatial dependence and a functional covariate}
\runtitle{GFLM under spatial dependence}

\begin{aug}
\author[A]{\fnms{Sooran}~\snm{Kim}\ead[label=e1]{Sooran.Kim@nyulangone.org}}, 
\author[B]{\fnms{Mark}~\snm{Kaiser}\ead[label=e2]{mskaiser@iastate.edu}}
\and
\author[C]{\fnms{Xiongtao}~\snm{Dai}\ead[label=e3]{xdai@berkeley.edu}}

\address[A]{Division of Biostatistics,
New York University\printead[presep={,\ }]{e1}}

\address[B]{Department of Statistics, 
Iowa State University\printead[presep={,\ }]{e2}}

\address[C]{Division of Biostatistics, 
	University of California\printead[presep={,\ }]{e3}}

\runauthor{S. Kim, M. Kaiser, and X. Dai}
\end{aug}

\begin{abstract}
We extend generalized functional linear models under independence to a situation in which a functional covariate is related to a scalar response variable that exhibits spatial dependence---a complex yet prevalent phenomenon.
For estimation, we apply basis expansion and truncation for dimension reduction of the covariate process followed by a composite likelihood estimating equation to handle the spatial dependency. We establish asymptotic results for the proposed model under a repeating lattice asymptotic context, allowing us to construct a confidence interval for the spatial dependence parameter and a confidence band for the regression parameter function. 
A binary conditionals model with functional covariates is presented as a concrete illustration and is used in simulation studies to verify the applicability of the asymptotic inferential results. 
We apply the proposed model to a  problem in which the objective is to relate annual corn yield in counties of states in the Midwestern United States to daily maximum temperatures from April to September in those same geographic regions. 
The extension to an expanding lattice context is further discussed in the supplement.
\end{abstract}



\end{frontmatter}

\section{Introduction}

In recent years, functional data analysis (FDA) has seen rapid development; see, for example, \cite{RS05}, \cite{HK12}, \cite{HE15}, and \cite{KR17}.
Functional regression models, in particular, have received attention due to their widespread applicability, with a comprehensive review available in \cite{M15}. Our specific interest lies in generalized functional linear regression models (GFLMs), which are models with functional covariates and potentially non-Gaussian real-valued responses and nonlinear expectation functions.
Previous works on GFLMs include \cite{J02}, \cite{JS05}, \cite{MS05}, \cite{CSP05}, and \cite{G11}. 
A challenge in handling functional data is the infinite dimensionality, a concern addressed, for instance, through a truncation strategy for dimension reduction of the covariate process \citep[cf.][]{MS05}.

In spatial statistics, functional covariate processes have also been considered, where a recent overview is given in \cite{M20}. 
Often, we may have interest in a given spatially structured response at a single point in time, but believe that the response is influenced by some historical covariate process evolving over time at individual spatial locations. Such problems can occur in fields like meteorology, ecological and environmental sciences, and social sciences, where some type of an event or response is influenced by environmental or behavioral conditions that develop over a prior time span.  In some problems, the response variables may exhibit spatial structure beyond that induced by the covariate process.   
For example, a scientific question of interest is the effect of the temporal trajectory of maximum daily temperatures on corn yield, but yield may exhibit spatial patterns due to other effects as well, such as soil fertility or rainfall gradients. 
Our goal then is to develop GFLMs that also incorporate direct spatial dependence among response variables.

Previous investigations have covered certain aspects of this setting, but fall short of capturing the full scope of our proposed model.
For instance, \cite{W11} studied asymptotic theory for generalized estimating equation {estimators} for binary responses with high-dimensional covariates, rather than a functional covariates; \cite{JKL17} suggested a GFLM for a functional covariate and real-valued longitudinal responses that have dependence over time; \cite{MS20} proposed a practical modeling approach for spatial lattice correlated data using a generalized estimating equation with Moran's index, considering spatial dependence but lacking a functional covariate and not investigating asymptotic properties.
In bridging this gap, we introduce a novel generalized linear model, which integrates both spatial dependence and functional covariates. 
In our model, responses given covariates follow (conditional) exponential dispersion family distributions, while the covariate processes are functional.
This innovative approach is built on a backbone of Markov random field models for spatially dependent responses and the truncation strategy 
applied to functional covariates.

We propose the use of composite likelihood for estimation and inference, a method commonly employed in spatial statistics but less typical in FDA. Through maximum composite likelihood estimation, we develop asymptotic results that represent a significant advancement in merging spatial statistics and FDA.
It is important to note that the basis coefficients and eigenvalues must decay to zero in an infinite-dimensional space, which is a distinct requirement that is not met by certain technical assumptions found in high-dimensional literature \citep[e.g.,][]{W11}. 
To overcome this limitation, we apply an alternative condition and provide new proof in this work. 
Our methodology is further validated through simulation studies, wherein our proposed model and method, which embrace spatial dependence, outperform an existing method for non-spatial GFLMs.

Our proposal holds significant value for analyzing complex real datasets characterized by both spatial dependence and the infinite dimensionality of covariate processes. For instance, these datasets may include temperature and corn yield data as previously mentioned, where we observe maximum temperatures from April to September over ten years alongside annual corn yield over the same period.
In this dataset, we encounter actual replications spanning ten years of spatial random fields, aligning with what is called a \textit{repeating lattice} asymptotic context in the development of inferential procedures.
We extend our discussion to situations where analyses must be conducted using only one observed spatial field, termed the \textit{expanding lattice} asymptotic context for inference. In this context, we examine COVID-19 data in 2021 to determine whether there is evidence of a relationship between the temporal progression of disease prevalence and subsequent willingness to receive full vaccination. The extension to expanding lattice context can be found in the supplement.

Our work introduces several novel contributions to the literature.
To the best of our knowledge, this is the first attempt to incorporate spatial dependence into functional linear regression with non-Gaussian response, while also providing asymptotic results. This work serves as a stepping stone for integrating spatial dependence within a GFLM framework.
Although composite likelihood methods are widely used in spatial analysis, applying them in the context of FDA is a novel contribution that expands the methodology in the FDA field.
Our theoretical contributions include establishing the consistency of the maximum composite likelihood estimator (MCLE). Such consistency is a fundamental result to derive the asymptotic normality of the estimator, which was implicitly assumed in \cite{MS05} for their quasi-likelihood estimator.
Additionally, we provide the asymptotic normality of quadratic forms of the MCLE by addressing challenges that arise from the sandwich information matrix, also known as the Godambe information. The proposed method is numerically demonstrated through simulation studies, and its practical utility is illustrated using a data example.

The rest of this article is organized as follows. 
In \autoref{s2}, we outline the background of the Markov random field approach and propose the spatial generalized linear model with a functional covariate.
We present an estimation strategy using composite likelihood in \autoref{s3}, and give theoretical results in \autoref{s4}.
Results from a simulation study are provided in \autoref{s5}.
\autoref{s6} contains the practical applications of our model, covering applications to maximum temperatures and corn yield. 
Concluding remarks are given in \autoref{s7}.
The proofs of the main theorems 
and the extension to expanding lattice context 
are provided in Section~S1 and Section~S5 of the supplement, respectively.

\section{Model}\label{s2}
We start with a review of Markov random field models, which form the foundation for our proposed model, in \autoref{s2.1}.
Our spatial generalized functional linear model (SGFLM) is then introduced in \autoref{s2.2}, followed by a specific example of our model in \autoref{s2.3}.

\subsection{Review of Markov Random Fields}\label{s2.1}
We begin this section with a brief overview of exponential family Markov random field models in the spatial context.
Let $\mbs_i$ denote spatial locations on a finite index random field, for $i=1, \ldots, n$.  For example, we may have $\mbs_i=(u_i,v_i)$, where $u_i$ denotes longitude or a horizontal coordinate on a regular lattice, and $v_i$ denotes latitude or a vertical coordinate on a lattice.  
Let $\{Y(\mbs_i): i= 1, \ldots, n \}$ be random variables indexed to the spatial locations and let $f$ be generic notation for a probability density (or mass) function such that $f(x)$ is the density of the random variable $X$, $f(y)$ is the density of the random variable $Y$, and so forth, where such variables may be univariate or multivariate.
Distributions of the  univariate $Y(\mbs_i)$ are assigned as full conditional distributions, $f(y(\mbs_i)| \{ y(\mbs_j): \, j \neq i\})$.
Now, for each location $\mbs_i$, we define a  neighborhood $N_i$ as
\begin{align*}
	N_i = \{\mbs_k: \,  f(y(\mbs_i)| \{ y(\mbs_j): \, j \neq i\}) \text{ depends functionally on } y(\mbs_k) \text{ for } k \neq i \},
\end{align*}
and $\mby(N_i) = \{ \mby(\mbs_j): \, \mbs_j \in N_i \}$ denotes the collection of the variables in the neighborhood $N_i$.
Similar to the Markov assumption in time, the Markov assumption in Markov random fields posits that the full conditional distribution is equivalent to the distribution conditioned only on neighboring variables,
\begin{displaymath}
	f(y(\mbs_i) |  \{ y(\mbs_j): \, j \neq i\}) = f(y(\mbs_i) | \mby(N_i) ).
\end{displaymath}

Typically, neighborhood structures are designated as a part of model formulation.  
Common examples of neighborhood structures on a regular lattice include the four-nearest and eight-nearest neighborhood structures.
If $\mbs_i=(u_i, v_i)$ where $u_i$ and $v_i$ are horizontal and vertical coordinates, respectively, the four-nearest neighborhood is given as,
\begin{displaymath}
	N_{4,i} = \{ \mbs_j: \, \, (u_j = u_i, v_j = v_i \pm 1) \text{ or } (u_j = u_i \pm 1, v_j = v_i) \}, 
\end{displaymath}
and an eight-nearest neighborhood is given as,
\begin{displaymath}
	N_i = N_{4,i} \cup \{ u_j = u_i\pm 1, \, v_j = v_i \pm 1 \}.
\end{displaymath}

Within the context of generalized linear models, the full conditional distributions in a Markov random field model are specified as one-parameter exponential families \citep[cf. ][]{B74},
\begin{equation} \label{condist}
	f(y(\mbs_i)|\mby(N_i), \bm{\theta}) = \exp\left[
	A_i(\mby(N_i), \bm{\theta}) y(\mbs_i) - B_i(\mby(N_i), \bm{\theta}) + C(y(\mbs_i))
	\right]
\end{equation}
where $A_i(\mby(N_i))$ is a natural parameter function that depends on the neighboring values of $Y(\mbs_i)$, and $B_i(\mby(N_i))$ is a function of $A_i(\mby(N_i))$ that determines the distributional form and moments. The function $C(\cdot)$ normalizes the distribution.
Now let $\mbtheta$ denote any parameters that may appear in the specified form of the full conditional distributions.  
Under Assumptions~1-2 in \cite{B74}, the author showed that the natural parameter function $A_i(\cdot)$ must satisfy
$A_i(\mby(N_i),\mbtheta) = \alpha_i + \sum_{\mbs_j \in N_i} \eta_{i,j} y(\mbs_j)$
where $\alpha_i$ is a leading constant and 
$\{\eta_{i,j} \}$ are dependence parameters satisfying $\eta_{i,j}=\eta_{j,i}$, $\eta_{i,i}=0$, and $\eta_{i,j}=0$ unless $\mbs_j \in N_i$. 
If external covariates influence the conditional distributions (\ref{condist}), they are incorporated into the leading term $\alpha_i$ using, for example, a standard link function from generalized linear models. To improve the interpretability of regression parameters in such a model,   \cite{KCF12} proposed to instead consider a centered parameterization,
$A_i(y(\mbs_i),\mbtheta) = \tau^{-1}(\kappa_i) + \sum_{\mbs_j \in N_i}\eta_{i,j}\{y(\mbs_j)-\kappa_{j}\}$,
where 
$\tau^{-1}(\cdot)$ is a function that maps expected values into natural parameters, and
$\kappa_i$ is the expected value of $Y(\mbs_i)$ under an independence model.
It is worth noting that if the dependence parameters $\{\eta_{i,j}\}$ become too large in magnitude, the model can suffer from degenerate behavior in which only a few elements in the joint support have non-negligible probability \citep[cf. ][]{KCF12}. 

\subsection{Spatial Generalized Functional Linear Model (SGFLM)}\label{s2.2}
In this section, we propose conditional exponential family distributions that involve a functional covariate.
Let $\kll_2(\ttt)$ be the set of all square-integrable functions on a closed interval $\ttt$. 
A functional covariate $X_i$ is a random element that takes values in $\kll_2(\ttt)$ with zero mean $\eo X_i=0$ for $i \in \{1,\cdots, n\}$.
Under constant dependence, that is, $\eta_{i,j}=\eta$ for all $i$ and $j$ that are neighbors,
a spatial generalized linear model with functional covariates and real-valued responses $y(\mbs_i)$ is, for $i=1, \ldots, n$,
\begin{equation}\label{model}
	\begin{split}
		f(y(\mbs_i)|\mby(N_i), X_i, \mbtheta) &= \exp\left[
		A_i(\mby(N_i), \mbtheta) y(\mbs_i) - B_i(\mby(N_i), \mbtheta) + C(y(\mbs_i))
		\right],
		\\
		A_i(\mby(N_i), \mbtheta) &= \tau^{-1}(\kappa_i) + \eta \sum_{\mbs_j \in N_i} \{y(\mbs_j)-\kappa_{j} \},	
		\\
		B_i(\cdot, \mbtheta) &\text{ is a function of }A_i(\cdot, \mbtheta), 	
		\\
		g(\kappa_i) &= \alpha + \int \beta(t) X_i(t) \, dt.   
	\end{split}
\end{equation}
Here, $\eta$ is a spatial dependence parameter,
$\tau^{-1}(\cdot)$ is a  function that maps expected values into natural parameters for the desired exponential family,
and $g(\cdot)$ is a link function that relates expectations under independence, $\kappa_i$, 
to a linear predictor with  intercept parameter $\alpha$, and a parameter function $\beta$ which is assumed to be in $\kll_2(\ttt)$.  The case in which $g(\kappa_i) = \tau^{-1}(\kappa_i)$ corresponds to a model with canonical link, and if $\eta =0$ model (\ref{model}) reduces to the GFLM of \cite{MS05}.
In this model, the large-scale structure of the spatial model for responses, namely the mean structure of independence model, is modeled via $\kappa_i$ through a functional covariate; 
and the small-scale structure, namely the spatial dependency, is modeled through the spatial dependence parameter $\eta$. Note that we have written the random model component in (\ref{model}) without the common dispersion parameter in the generalized linear model.  A separate dispersion parameter could be added when needed, such as for Gaussian or Inverse Gaussian random components.  The most common non-Gaussian models in spatial applications are natural exponential families such as binary and (Winsorized) Poisson models for which the dispersion parameter can be taken as $1$.  

A basis expansion can be employed to handle functional covariates in model (\ref{model}).
Let $\{\phi_j\}_{j=1}^\infty$ be an orthonormal basis of the functional space $\kll_2(\ttt)$, for example, a trignometric basis. 
A basis expansion can be used to get the Fourier series of the functional covariate $X_i$ and the parameter function $\beta$ as
$$X_i = \sum_{j=1}^\infty \varepsilon_j^{(i)} \phi_j , \quad \beta = \sum_{j=1}^\infty \beta_j \phi_j,$$
where $\varepsilon_j^{(i)} = \int X_i(t) \phi_j(t) dt$ and $\beta_j = \int \beta(t) \phi_j(t) dt$ denote their Fourier coefficients.
The implication is  that the inner product in (\ref{model}) can be expressed with the infinite sum of their Fourier coefficients, that is,
$$\int \beta(t)X_i(t) dt =  \sum_{j=1}^\infty \beta_j \varepsilon_j^{(i)}.$$
As a result, the systematic model component in model (\ref{model}) can be written as
\begin{align}\label{inf_sum}
	g(\kappa_i) &= \alpha + \sum_{j=1}^\infty \beta_j \varepsilon_j^{(i)}.
\end{align}	

We approximate the infinite sum in (\ref{inf_sum}) using a truncated version at $p$ terms,
following the truncation strategy \citep[cf.][]{MS05}. The $p$-truncated model is formed through a truncation of the basis expansion,
\begin{align}\label{p_model}
	g(\kappa_i) &= \alpha + \sum_{j=1}^{p} \beta_j \varepsilon_j^{(i)},
\end{align}
where asymptotically $p$ diverges as the independent realizations or the sample size diverges, depending on the asymptotic context.
With the $p$-truncation in (\ref{p_model}), we can consider the $(p+2)$-dimensional parameter vector $\mbtheta=(\eta, \alpha, \beta_1, \cdots, \beta_p)^\top$ instead of a parameter function.
This $p$-truncated model will be adopted for the estimation and asymptotic inference of the parameters $\eta$ and $\beta$, which are developed in Sections~\ref{s3}-\ref{s4}.

It is worth noting that our models (\ref{model}) or (\ref{p_model}) incorporate both a spatial dependence parameter $\eta$, and functional covariate $X$ along with parameter function $\beta$, or Fourier coefficient $\varepsilon_j$ with truncated parameter $\beta_j$ that increases asymptotically.
This is the first successful endeavor to investigate spatial dependence and functional covariate processes together in a GLM setting, building on the work of \cite{MS05}.

\subsection{Binary Conditional Model with Functional Covariates}\label{s2.3}
We now present a specific example of model~\eqref{model} to make the concepts and notation more concrete, that being a model with binary conditional distributions.  This model will also be used in the simulation study to follow, as well as the applications.	
	Suppose that we are interested in spatially dependent binary responses at locations $\{ \mbs_i: \, i=1, \ldots, n\}$ with a functional covariate $X_i$.
	For $\mbtheta=(\eta, \alpha, \beta_1, \cdots, \beta_p)^\top$, the $p$-truncated model can be written as,
	\begin{equation}\label{model:ber}
		\begin{split}
			f(y(\mbs_i)|\mby(N_i), X_i, \mbtheta) &= \exp\left[
			A_i(\mby(N_i), \mbtheta) y(\mbs_i) - B_i(\mby(N_i), \mbtheta)
			\right],
			\\
			A_i(\mby(N_i), \mbtheta) &= \log\left(\kappa_i \over 1-\kappa_i \right) + \eta \sum_{\mbs_j \in N_i} \{y(\mbs_j)-\kappa_j \},	
			\\
			B_i(\mby(N_i), \mbtheta) &=\log\left(1+e^{A_i\left(\mby(N_i), \mbtheta\right)}\right),	
			\\
			\log\left(\kappa_i \over 1-\kappa_i \right) &=\alpha + \sum_{j=1}^p \beta_j \varepsilon_j^{(i)}	
		\end{split}
	\end{equation}
	where $\varepsilon_j^{(i)} = \int X_i(t) \phi_j(t) dt$ and $\beta_j = \int \beta(t) \phi_j(t) dt$ 
	for an orthonormal basis $\{\phi_j\}_{j=1}^\infty$.

\section{Maximum Composite Likelihood Estimation}\label{s3}
This section provides an outline of the estimation method using composite likelihood.
In our model described in (\ref{model}), we specify full conditional distributions. 
Under assumptions in Theorem 3 of \cite{KC00}, a joint distribution corresponding to the set of full conditional distributions specified in the model exists, but aside from  models with Gaussian conditionals, that joint distribution has an intractable form. 
While our assumption of constant dependence (along with the positivity condition), satisfying $\eta_{i,j} = \eta$ for all $i$ and $j$, is sufficient to allow a joint distribution to be identified through the use of the negpotential function of \cite{B74} \citep[cf.][]{KC00}, it is difficult to use maximum likelihood estimation due to an intractable normalizing term. 
Hence, we propose to use composite likelihood for estimation, which is also common in purely spatial applications.

A composite likelihood is defined by multiplying a set of component likelihoods, selected based on the context of the problem. Several examples include pairwise marginal likelihood, pairwise conditional likelihood, pairwise difference likelihood, and Besag's original pseudo-likelihood; see \cite{V11} for more details. 
The composite likelihood we will make use of corresponds to the original pseudo-likelihood of \cite{B75}. 
Given a set of full conditional density or mass functions in the form of (\ref{model}), define the composite likelihood as,
$$L_c(\mbtheta|\bm{y})= \prod_{i=1}^n f(y(\mbs_i) | \mby(N_i), X_i, \mbtheta ).$$
In the case of our binary conditional model with functional covariates described in (\ref{model:ber}), 
for $\mbtheta=(\eta, \alpha,\beta_1, \cdots, \beta_p)^\top$, the log pseudo-likelihood is,
\begin{align*}
	l_c(\mbtheta|\bm{y})
	&=\sum_{i=1}^n \log f(y(\mbs_i) | \mby(N_i), X_i, \mbtheta)
	=\sum_{i=1}^n \left\{A_i(\mby(N_i),\mbtheta) y(\mbs_i) - B_i(\mby(N_i),\mbtheta)  \right\},
\end{align*}
where $A_i(\mby(N_i), \mbtheta)$ and  $B_i(\mby(N_i), \mbtheta)$ are defined in (\ref{model:ber}).
The maximum composite likelihood estimator (MCLE) $\hat{\mbtheta}$ can be obtained as
$\hat{\mbtheta} = \argmax l_c(\mbtheta|\bm{y})$,		
and then the estimated parameter function can be obtained as $\hat{\beta}=\sum_{j=1}^p \hat{\beta}_j \phi_j$.


One practical challenge lies in selecting initial values for optimization to obtain MCLE, which typically involves the use of iterative procedures.
In our implementation, the composite likelihood function is maximized using numerical optimization routines available in \texttt{R} (e.g., \texttt{optim}). 
Since the performance of such algorithms may depend on the choice of starting values, careful initialization is important to ensure stable convergence.
To obtain reasonable initial values, we proceed in a sequential manner.
First, we select the initial values for $\alpha$ and $\{\beta_j\}_{j=1}^p$, by temporarily assuming a fixed value of $\eta$.
These initial estimates can be obtained from functional linear regression models that ignore spatial dependence, such as the functional principal component regression approach described in \cite{H07}.
Next, an initial value for $\eta$ is obtained, while fixing $\alpha$ and $\beta_j$ at their preliminary estimates, using a log-likelihood slice method.
Specifically, the value that maximizes a log composite likelihood slice using a one-dimensional optimization algorithm, such as an equal interval search or bisection, is selected as an initial value for $\eta$.

In the situation where the expectation of the functional covariate is non-zero, $\eo X_i \neq 0$,  
we use the functional covariates $X_i^{\mathrm{center}} = X_i - \bar{X}$ centered by the sample mean $\bar{X}= n^{-1}\sum_{i=1}^n X_{i}$ instead of original covariates $X_i$.
When employing centered functional covariates, the estimated truncated parameter $\hat{\beta}_j^{\mathrm{center}}$ remains the same as the original estimate $\hat{\beta}_j$.
However, the estimated intercept parameter $\hat{\alpha}^{\mathrm{center}}$ should be adjusted from the original estimate $\hat{\alpha}$.
The adjusted estimate of the intercept parameter $\hat{\alpha}^{\mathrm{center}}$ is given by
$$\hat{\alpha}^{\mathrm{center}} = \hat{\alpha} + \sum_{j=1}^{p} \hat{\beta}_{j}^{\mathrm{center}} \int \bar{X}(t) \phi_j(t) dt.$$

\section{Asymptotic Inference}\label{s4}

There are two common types of asymptotic context in spatial statistics with discrete spatial indices, typically referred to as the {\it repeating lattice} context and the {\it expanding lattice} context \citep[cf.][]{V11}. The repeating lattice context refers to the sample size growing large through independent realizations of a fixed grid structure.   
It means that we have $N$ independent realizations $\{(\bm{X}_k, \bm{y}_k)\}_{k=1}^N$
where $\bm{X}_k=(X_{k,1}, \cdots, X_{k,n})^\top$ and $\bm{y}_k = (y(\mbs_1)_k, \cdots, y(\mbs_n)_k)^\top$ 
follow model (\ref{model}).
We refer to the number $N$ of independent realizations as the \textit{repeating size}.
In this asymptotic context, we can add log composite likelihoods of the independent realizations, similarly to what we would do in the usual independent and identically distributed (iid) case. 
We do assume that the truncation level $p=p_N$ increases asymptotically as the repeating size $N$ goes to infinity. 
Throughout this section, our focus is on the repeating lattice context, while we will discuss the extension to the expanding lattice context in the supplement.

Composite likelihood asymptotics under a repeating lattice context generally involve the Godambe information matrix \citep[cf.][]{G60}, which is defined as the
sandwich information matrix.
If the composite likelihood is a true log likelihood function, then the Godambe information reduces to  Fisher information. 
The precise definition of Godambde information will be presented in \autoref{s4.1}.
In the finite-dimensional case, under regularity conditions, we may expect an asymptotic normality with asymptotic variance being the inverse Godambe information as follows: 
\begin{align}\label{normality_spatial}
	N^{1/2} [G(\bm{\theta})]^{1/2} (\hat{\bm{\theta}}-\bm{\theta}) \cond \normal(\bm{0}_p, I_{p\times p})
\end{align}
where $\hat{\bm{\theta}} \in \R^p$ is the MCLE and $\normal(\bm{0}_p, I_{p\times p})$ denotes the $p$-dimensional normal distribution with mean $0 \in \R^p$ and covariance matrix equal to the $p\times p$ identity matrix $I_{p\times p}$ \citep[cf.][]{L88, V11}. 
However, in FDA, such weak convergence under strong norm may not be feasible even when the responses are independent \citep[cf.][]{CMS07}. Therefore, we instead derive an asymptotic normality for quadratic form \citep[cf.][]{MS05}.
The Godambe information matrix will play an important role in our inference, particularly related to the asymptotic variance. 

\autoref{s4.1} introduces the notations used in the theoretical development.
\autoref{s4.3} establishes the consistency of the maximum log composite likelihood estimator and provides limiting distribution results.
The technical conditions required for the asymptotic theory are presented in \autoref{s4.2}.

\subsection{Notation}\label{s4.1}
In this section, we introduce some notation that will be used in the sequel.
%
The Euclidean norm $\|\bm{\cdot}\|_2$ is defined by $\|\bm{x}\|_2=(x_1^2 + \cdots x_n^2)^{1/2}$ for $\bm{x}=(x_1, \cdots, x_n)^\top \in \R^n$.
Let $\ip{\cdot, \cdot}_F$ denote the Frobenius inner product between matrices, defined by $\ip{A,B}_F = \tr(A^\top B)$ for $m \times n$  matrices $A$ and $B$,
which induces the Frobenius norm $\|\cdot \|_F$ as $\|A\|_F=(\sum_{i=1}^m \sum_{j=1}^n a_{ij}^2)^{1/2}$ for a matrix $A$.

Let $l_c(\bm{\theta}|\bm{y}_k)$ be the log composite likelihood of $\bm{y}_k$
for a $(p_N+2)$-dimensional parameter $\bm{\theta} = (\eta, \alpha, \beta_1, \cdots, \beta_{p_N})^\top$.
In the rest of the paper, let $\alpha = \beta_0$ following \cite{MS05}. 
We list some additional notations related to the log composite likelihood, such as its derivatives and their empirical averages:
$l_{c,N}(\bm{\theta})=\sum_{k=1}^N l_c(\bm{\theta}|\bm{y}_k)$,
$\bar{l}_{c,N}(\bm{\theta})=N^{-1} l_{c,N}(\bm{\theta})$, 
$\bar{l}_{c,N}'(\bm{\theta})=N^{-1} l'_{c,N}(\bm{\theta})$, 
$\bar{l}_{c,N}''(\bm{\theta})=N^{-1} l''_{c,N}(\bm{\theta})$, and 
$\bar{l}^{(3)}_{c,N}(\bm{\theta})\\=N^{-1} l^{(3)}_{c,N}(\bm{\theta})$.
Let $H(\bm{\theta}) = \eo[-l_c''(\bm{\theta}|\bm{y}_1)]$ and $J(\bm{\theta}) = \eo[l_c'(\bm{\theta}|\bm{y}_1) l_c'(\bm{\theta}|\bm{y}_1)^\top] = [j_{uv}]_{u,v=1}^{p_N+2}$ respectively denote the expected value of the negative second derivative $l_c''(\bm{\theta}|\bm{y}_1)$ and covariance matrix of the first derivative $l_c'(\bm{\theta}|\bm{y}_1)$.
We assume that their inverses $H(\bm{\theta})^{-1} = [h^{-1}_{ij}]_{1 \leq i,j \leq 2}$ and 
$ J(\bm{\theta})^{-1} = W(\bm{\theta}) = [w_{uv}]_{u,v=1}^{p_N+2}$ exist, 
where $h_{11}^{-1} \in \R$ and a $(p_N+1)\times (p_N+1)$ matrix $h_{22}^{-1}$ denote block diagonal elements of $H(\bm{\theta})^{-1}$.

The Godambe information matrix $G(\bm{\theta})$, which is crucial for the asymptotic variance, is defined as $G(\bm{\theta}) = H(\bm{\theta})J(\bm{\theta})^{-1} H(\bm{\theta})=[G_{ij}(\bm{\theta})]_{1 \leq i,j \leq 2}$, where $G_{11}(\bm{\theta}) \in \R$ and $(p_N+1) \times (p_N+1)$ matrix $G_{22}(\bm{\theta})$ are its block diagonal elements. 
Positive definiteness is presumed for the Godambe information.
Its inverse $G(\bm{\theta})^{-1}$ is decomposed as $G(\bm{\theta})^{-1} = H(\bm{\theta})^{-1} J(\bm{\theta}) H(\bm{\theta})^{-1} = [G_{ij}^{(-1)}(\bm{\theta})]_{1 \leq i,j \leq 2}$; we use a block matrix representation to $G(\bm\theta)^{-1}$ with $G_{11}^{(-1)}(\bm{\theta}) \in \R$ and a $(p_N+1) \times (p_N+1)$ matrix $G_{22}^{(-1)}(\bm{\theta})$.
Both the Godambe information matrix $G(\bm{\theta})$ and its inverse $G(\bm{\theta})^{-1}$ are well-defined due to the invertibility of $J(\bm{\theta})$ and $H(\bm{\theta})$.

Lastly, 
we introduce sequences $a_N$, $b_N$, and $c_N$ of positive real numbers defined as follows:
\begin{align}\label{abc}
	a_N =\|H(\bm{\theta})^{-1}\|_F^2,
	\quad
	b_N = \ld_{\textrm{min}}(H(\bm{\theta})),
	\quad
	c_N=\|J(\bm{\theta})^{-1}\|_F^2.
\end{align}
The sequences $a_N$ and $c_N$ represent the squared Frobenius norm of the inverses $H(\bm{\theta})^{-1}$ and $J(\bm{\theta})^{-1}$, respectively,
while $b_N$ denotes the minimum eigenvalue of $H(\bm{\theta})$, which may not be bounded away from zero as opposed to high-dimensional literature \citep[cf.][]{W11}.
%
%
These norms of inverses, which depend on the increasing dimension $p_N$, and the minimum eigenvalue which goes to zero should be involved theoretically, particularly when focusing on the truncated model. They play an important role in our theoretical framework.


\reqnomode

\subsection{Main Results}\label{s4.3}
The assumptions used in the main theorems are detailed in \autoref{s4.2}.
We first establish the existence and consistency of the MCLE $\hat{\bm{\theta}}$, followed by the asymptotic normality of the quadratic form involving the MCLE $\hat{\bm{\theta}}$. 
Lastly, we provide the asymptotic results for the separate parameters: the spatial dependence parameter $\eta$, and slope coefficient vector $\bm{\beta}=[\beta_0, \beta_1, \cdots, \beta_{p_N}]^\top$. 
From these results, confidence sets for dependence parameter $\eta$ and regression parameter function $\beta$ can be constructed distinctly.
The proofs of the main theorems 
are provided in Sections~S1 of the supplement.

To discuss asymptotic normality results in Theorems~\ref{normality_theta}-\ref{normality_eta}, the existence and consistency of $\hat{\bm{\theta}}$ are crucial. We provide the existence and consistency of $\hat{\bm{\theta}}$, which is the most fundamental result, in the following theorem.
\begin{thm}\label{cons}
	Suppose Assumptions (A1)-(A3) 
	hold along with $N^{-1/2} p_N b_N^{-2}  \rightarrow 0$ as $N \rightarrow \infty$. 
	Then, there exists a solution $\hat{\bm{\theta}}$ of the equation $l_{c,N}'(\bm{\theta})=0$ that is consistent for $\bm{\theta}$, in the sense that $\|\hat{\bm{\theta}} -\bm{\theta}\|_2 \to 0$ in probability as $N\to\infty$. Specifically, it satisfies that
	$$\|\hat{\bm{\theta}} -\bm{\theta}\|_2=O_p\left(N^{-1/2} p_N^{1/2} b_N^{-1}\right).$$
\end{thm}
In a high-dimensional case, a convergence rate may solely depend on $N$ and $p_N$, as exemplified by Theorem~3.6 of \cite{W11}, which provides a convergence rate of $O_p(N^{-1/2} p_N^{1/2})$. 
However, functional covariates may lie in an infinite-dimensional space. It suggests the essential role of the eigendecay represented by $b_N$ in determining the convergence rate in \autoref{cons}, in contrast to high-dimensional covariates.

\begin{rem}\label{remark1}
	We assume a polynomial decay rate for $b_N$, defined in \eqref{abc}, given by $b_N \asymp p_N^{-(1+\gamma)}$ and the polynomial growth rate for $p_N \asymp N^\rho$, where $\gamma, \rho \in (0,\infty)$.
	Under these assumptions, the condition $N^{-1/2} p_N b_N^{-2}  \rightarrow 0$ 
	holds if $\rho \in (0, 1 / (4\gamma+6))$,
	since $N^{-1/2} p_N b_N^{-2} \asymp N^{-1/2}p_N^{2\gamma+3} \asymp N^{\rho(2\gamma+3)-1/2}$. 
	For example, if $\gamma=1/5$ and $\rho = 1/7$, this condition holds with a subsequent convergence rate of $N^{-9/35}$. 
	In the same setup, 
	the upper bound for the rate established by \cite{W11} in high-dimensional literature is $N^{-1/2} p_N^{1/2} \asymp (N^{-1/2})^{1-\rho} = N^{-3/7}$, which is faster than $N^{-9/35}$. 
	This is expected, as eigenvalues are anticipated to decay to zero with functional data, while high-dimensional literature often considers a fixed lower bound for the eigenvalues.
\end{rem}

Given the infinite dimensionality inherent of functional data, obtaining asymptotic results directly for $\hat{\bm{\theta}}$ may be challenging. Hence, we rely on the asymptotic normality of its quadratic form, which is given next.
\begin{thm}\label{normality_theta}
	Recall that $a_N$, $b_N$, and $c_N$ are defined in \eqref{abc}.
	Suppose Assumptions (A1)-(A7) 
	hold along with $N^{-1/2}p_N a_N  b_N^{-2}\rightarrow 0$, and $N^{-1/3} p_N \rightarrow 0$ as $N \rightarrow \infty$. Then, as $N \rightarrow \infty$,
	$${N (\hat{\bm{\theta}}-\bm{\theta})^\top G(\bm{\theta}) (\hat{\bm{\theta}}-\bm{\theta}) - (p_N+2) \over \sqrt{2(p_N+2) }} \cond \normal(0,1).$$
	If additionally Assumption (G1) 
	holds along with
	$N^{-1} p_N^7 b_N^{-6}  c_N^2 \rightarrow 0$ as $N \rightarrow \infty$, 
	then the above asymptotic normality remains valid replacing the Godambe information $G(\bm{\theta})$ by its empirical counterpart $\hat{G}_N(\hat{\bm{\theta}}) = \hat{H}_N(\hat{\bm{\theta}}) \hat{J}_N(\hat{\bm{\theta}})^{-1} \hat{H}_N(\hat{\bm{\theta}})$,
	where
	$\hat{H}_N(\hat{\bm{\theta}}) = N^{-1} \sum_{k=1}^N (-l_c''(\hat{\bm{\theta}}|\bm{y}_k))$ and
	$\hat{J}_N(\hat{\bm{\theta}}) = N^{-1} \sum_{k=1}^N l_c'(\hat{\bm{\theta}}|\bm{y}_k) l_c'(\hat{\bm{\theta}}|\bm{y}_k)^\top$ are respectively the empirical counterparts of $H(\bm{\theta})$ and $J(\bm{\theta})$.
\end{thm}
\begin{rem}\label{remark2}
	Similar to \autoref{remark1}, we assume that $p_N \asymp N^\rho$ and $b_N \asymp p_N^{-(1+\gamma)}$ for $\rho, \gamma \in (0,\infty)$. 
	Since $a_N \leq p_Nb_N^{-2}$, we obtain 
	$$N^{-1/2} p_N a_N b_N^{-2} \leq  N^{-1/2} p_N^2 b_N^{-4}  \asymp N^{-1/2} p_N^{4\gamma+6} \to 0$$ 
	if $\rho \in (0,1/(8\gamma+12))$. 
	Now, let $d_N = \lambda_{\min}(J(\theta))$ represent the minimum eigenvalue of $J(\theta)$, and suppose that $d_N \asymp b_N$. 
	Since $c_N \leq p_Nd_N^{-2}$, we also have 
	$$N^{-1} p_N^7 b_N^{-6} c_N^2 \leq N^{-1} p_N^{9} b_N^{-6} d_N^{-4} \asymp N^{-1} p_N^{10\gamma+19}\to 0$$ 
	if $\rho \in (0,1/(10\gamma+19))$. 
	For instance, 
	if $\gamma = 1/9$ and $\rho = 1/13$, only the first condition holds,
	while both conditions are satisfied when $\gamma = 1/11$ and $\rho = 1/20$.
	In the latter case, asymptotic normality in \autoref{normality_theta} holds with the empirical Godambe information $\hat{G}_N(\hat{\bm{\theta}})$.
	The growth rates of $p_N$ are quite slower compared to other previous work on functional regression, such as \cite{H07} and \cite{MS05}, under independence. 
	For example, in \cite{MS05}, a sufficient condition for similar asymptotic normality is $N^{-1/2}p_N^2 \to 0$, which holds for $\rho  \in (0, 1/4)$. 
	This suggests that inference under spatial dependence may require slower growth rates for $p_N$ compared to the independence case.
	Nevertheless, faster growth rates for $p_N$ could still be achievable  by employing more refined upper bounds for $a_N$ and $c_N$, as we have used rather conservative bounds, namely $a_N \leq p_Nb_N^{-2}$ and $c_N \leq p_Nd_N^{-2}$ in this discussion.
\end{rem}

We might be more interested in the inference of the slope coefficient vector $\bm{\beta}=[\beta_0, \beta_1, \cdots, \beta_{p_N}]^\top$ and the spatial dependence parameter $\eta$, as they can offer more practical use compared to the entire parameter $\bm{\theta}$. We furnish two distinct asymptotic normality results for both parameters, starting with the slope coefficient vector $\bm{\beta}=[\beta_0, \beta_1, \cdots, \beta_{p_N}]^\top$. 
Let $$\hat{G}_N(\hat{\bm{\theta}})^{-1} = \hat{H}_N(\hat{\bm{\theta}})^{-1} \hat{J}_N(\hat{\bm{\theta}}) \hat{H}_N(\hat{\bm{\theta}})^{-1} 
=\begin{bmatrix}
	\hat{G}_{11}^{(-1)}(\hat{\bm{\theta}}) & \hat{G}_{12}^{(-1)}(\hat{\bm{\theta}})
	\\ \hat{G}_{21}^{(-1)}(\hat{\bm{\theta}}) & \hat{G}_{22}^{(-1)}(\hat{\bm{\theta}})
\end{bmatrix}$$
be the inverse of the sample Godambe information.
The asymptotic result for the quadratic form of $\hat{\bm{\beta}}-\bm{\beta}$ can be derived as shown in the following theorem.
In FDA, a direct central limit theorem for the functional parameter estimator $\hat{\beta}$ is generally not available due to the infinite-dimensional nature of the parameter space \citep[cf.][]{CMS07}. Therefore, inference is often based on univariate summary statistics such as quadratic forms \citep[cf.][]{MS05, horvath2013estimation, kong2016classical}, which motivates the use of this quadratic form.
\begin{thm}\label{normality_beta}
	Suppose that Assumptions (A1)-(A5), (B1)-(B3) 
	hold along with $N^{-1/3} p_N \rightarrow 0$ and $N^{-1/2}p_N a_N  b_N^{-2}  \rightarrow 0$ as $N \rightarrow \infty$.
	We have, as $N \rightarrow \infty$,
	$${N (\hat{\bm{\beta}}-\bm{\beta})^\top (G_{22}^{(-1)}(\bm{\theta}))^{-1} (\hat{\bm{\beta}}-\bm{\beta}) - (p_N+1)  \over \sqrt{2 (p_N+1)}} \cond \normal(0,1).$$
	If additionally Assumptions (G1)-(G2) 
	hold along with 
	$N^{-1} p_N^7 b_N^{-6} c_N^2  \rightarrow 0$ as $N \rightarrow \infty$,		
	then the above asymptotic normality for $\hat{\bm{\beta}}$ remains valid replacing $(G_{22}^{(-1)}(\bm{\theta}))^{-1}$ by $(\hat{G}^{(-1)}_{22}(\hat{\bm{\theta}}))^{-1}$.
	Here, $a_N$, $b_N$, and $c_N$ are as defined in \eqref{abc}.
\end{thm}

It is worth noting that the quadratic form of $\hat{\bm{\beta}}-\bm{\beta}$ is still related to the entire parameter vector $\bm{\theta}$ through $G_{22}^{(-1)}(\bm{\theta})$, despite our focus on inference of $\bm{\beta}$. Nevertheless, \autoref{normality_beta} mirrors a result that bears similarity to Theorem 4.1 in \cite{MS05}.
Consequently, we can construct confidence bands of $\beta$ in a manner analogous to Corollary 4.3 in \cite{MS05}.
We formalize the confidence band for $\beta$ in the subsequent corollary.
\begin{cor}\label{conf_band} 
	We define $\bm{\phi}:\ttt\to\R^{p+1}$ by
	$\bm{\phi}(t)= [\phi_1(t), \cdots, \phi_{p+1}(t)]^\top$ for $t \in \ttt$. 		
	Under the assumptions of \autoref{normality_beta}, for large $N$ and $p_N$, an approximate $(1-\alpha)$ simultaneous confidence band is constructed by
	\begin{align*}
		\hat{\beta}(t) \pm \sqrt{c(\alpha) \bm{\phi}(t)^\top G_{22}^{(-1)}(\bm{\theta})  \bm{\phi}(t)},
	\end{align*}
	where $c(\alpha) = [(p+1+\sqrt{2(p+1)} \Phi(1-\alpha))]/N$ and
	$\Phi$ represents the cumulative distribution function of the standard normal distribution $\normal(0,1)$.
\end{cor}

In practical applications, these bands can be derived from the empirical counterpart of $G(\bm{\theta})$, denoted by $\hat{G}_N(\hat{\bm{\theta}})$. The proof of \autoref{conf_band} follows from the result presented in \autoref{normality_beta} and utilizes the same argument as in Corollary 4.3 of \cite{MS05}.

One of the main distinctions from the iid case is the inclusion of the spatial dependence parameter $\eta$. 
Lastly, we establish the asymptotic normality of the spatial parameter $\hat{\eta}$, which in turn facilitates the construction of confidence intervals for $\eta$, despite its connection to the entire parameter vector $\bm{\theta}$.
\begin{thm}\label{normality_eta} 
	Suppose that Assumptions (A1)-(A5), and (E1) 
	hold along with $N^{-1/2} p_N a_N  b_N^{-2} \rightarrow 0$ as $N \rightarrow \infty$,
	where $a_N$ and $b_N$ are defined in \eqref{abc}.
	Then, as $N \rightarrow \infty$,
	$$\sqrt{N} (G_{11}^{(-1)}(\bm{\theta}))^{-1/2} (\hat{\eta}-\eta)  \cond \normal(0,1).$$	
	If additionally Assumptions (G1) hold along with
	$N^{-1} p_N^5 a_N^3 b_N^{-4}  \rightarrow 0$ as $N\rightarrow \infty$,
	then the above asymptotic normality for $\hat{\eta}$ remains valid replacing $(G_{11}^{(-1)}(\bm{\theta}))^{-1/2}$ by $(\hat{G}_{11}^{(-1)}(\hat{\bm{\theta}}))^{-1/2}$.
\end{thm}

To conclude this section, we highlight the key theoretical difference from the work of \cite{MS05}. 
First, while the previous work implicitly assumes consistency, we make a deliberate effort to provide a formal proof in \autoref{cons}. 
More importantly, our results include a spatial dependence parameter, which is absent in the iid case.
Although the Godambe information cannot be decomposed into separate forms for $\bm{\beta}$ and $\eta$, which adds significant complexity to our scenario, we derive distinct normalities for the slope coefficient and spatial dependence parameter. This facilitates the construction of confidence bands or intervals easily.

\section{Simulation}\label{s5}
In this section, we discuss a Monte Carlo study conducted to assess the performance of our model, SGFLM, when estimated using composite likelihood.  
Since no existing methods account for spatial dependence in responses with a functional covariate process,
we contrast results for SFGLM with those for the non-spatial GFLM using quasi-likelihood estimation, following \cite{MS05}. 

We computed a number of mean squared errors for individual estimators of parameters, looked at coverage of approximate confidence intervals for the spatial dependence parameter, and compared a measure of goodness of fit between models that did and did not include spatial structure.  For the Monte Carlo results, we used $M=1,000$ simulated cases. 
Because our theoretical results relate to a context of repeating lattices, one simulated case consisted of $20$ data sets simulated from model (\ref{model:ber}) with binary responses,  $\{(\bm{X}_k, \, \bm{y}_k) \}_{k=1}^{20}$.  Each $\bm{X}_k$ and $\bm{y}_k$ contained values on a $20 \times 20$ regular lattice wrapped on a torus and using  a four-nearest neighborhood structure, so that $n = 400$ for each of the $20$ data sets in each Monte Carlo case.       

To simulate the functional regression model, we used the following strategy.  
Let  $\{\phi_j\}_{j=1}^{20}$ be the first 20 functions from the trigonometric base. 
To obtain independent copies of functional covariates $\bm{X}_k = \{ X_i: \, i=1, \ldots, n\}_k$ for $k = 1, \cdots, 20$, each functional covariate 
$X_i$ was generated as
$X_i=\mu + \sum_{j=1}^{20} \e_j^{(i)} \phi_j$
where $\mu(t)=4t \sin(3t)$ and $\e_j^{(i)} \sim N(0,1/j^2)$.  
Now take $\alpha = 0$ in (\ref{model:ber}) and let 
$\beta=\sum_{j=1}^{20} \beta_j \phi_j$, 
where $\beta_j=j^{-1}$ for $j=1,2,3$ and $0$ for $j>3$. 
All curves were produced at $50$ equally spaced values of $t$ between $0$ and $1$.

Simulation of response variables $\bm{y}_k = \{ y(\mbs_i): \, i=1, \ldots, n\}_k$ for $k=1, \ldots, 20$ was accomplished through the application of a Gibbs Sampling algorithm and the full conditional distributions from (\ref{model:ber}). 
In running the Gibbs algorithm, initial values for the $n=400$ spatial locations were generated from independent Bernoulli distributions with parameter $0.5$ and we set a burn-in period of $200$.  We then collected every $200th$ data set produced to obtain $20$ data sets for a case.  The overall algorithm was re-initialized for each of the $M=1,000$ cases.

In application, we need to select the truncation level $p$ in order to conduct estimation and inference.  
For this task, \cite{MS05} utilized Akaike Information Criterion (AIC) based on their simulation, which considered various criteria such as AIC, Bayesian Information Criterion (BIC), and minimization of the leave-one-out prediction error. 
We adopted an AIC-based criterion making use of the log composite likelihood, which was defined as,
\begin{align}\label{AIC}
	\text{AIC}_\text{c}=2(p+2)-2 l_{c,N}(\hat{\bm{\theta}}).
\end{align}

Performance measures were Monte Carlo approximations of expectation and mean squared error (MSE) for the scalar parameters $\eta$ and $\alpha$, defined for $\eta$ as,
$$\text{E}_M (\hat{\eta}) = {1 \over M} \sum_{m=1}^{M} \hat{\eta}_m \hspace*{1cm}   \text{MSE}_M(\hat{\eta})={1 \over M} \sum_{m=1}^M (\eta - \hat{\eta}_m)^2,$$
where $\hat{\eta}_m$ is the estimate of the spatial dependence parameter in $m$-th Monte Carlo case for $m=1,\cdots, M$,
and similarly for $\alpha$. For the parameter function $\beta$, we computed the mean integrated squared error (MISE) of a slope estimate defined as,
$$\text{MISE}_M(\hat{\beta}) = {1 \over M} \sum_{m=1}^M \int (\beta(t) - \hat{\beta}_m(t))^2 dt,$$
where $\hat{\beta}_m$ denotes the estimate of the parameter function in $m$-th Monte Carlo case for $m=1,\cdots, M$.
We also  computed a Monte Carlo approximation to  its integrated variance $\int \text{var}(\hat{\beta}(t))dt$ as,
$$\ \text{IV}_M(\hat{\beta}) = {1 \over M} \sum_{m=1}^M  \int (\hat{\beta}_m(t) - E[\hat{\beta}_m(t)])^2 dt.$$		
The empirical coverage of $95\%$ confidence intervals for $\eta$ and confidence bands for $\beta$ were obtained as,
\begin{align*}
	\text{CI}_M &= {1\over M} \sum_{m=1}^M \I(\eta \in \text{CI}_m),
	\\ \text{CB}_M &= {1\over M} \sum_{m=1}^M \I(\beta(t) \in \text{CB}_m(t) \text{ for all }t),
\end{align*}
respectively, where $\I$ is the indicator function, and
CI$_m$ and $\{\text{CB}_m(t):t \in [0,1]\}$ denote confidence interval and confidence band constructed from the $m$-th Monte Carlo case based on the results in \autoref{normality_eta} and \autoref{conf_band}, respectively. 
Finally, we used a fitted mean squared error (FMSE) criterion to compare the GFLM of \cite{MS05} which assumes spatial independence, and our proposed model (SGFLM) which incorporates spatial structure in the response variables.   This criterion is the average squared difference between observed and estimated conditional expected values and was computed as,
\begin{align}\label{FMSE}
	\text{FMSE}_M={1 \over M} \sum_{m=1}^M \left( {1\over N}   \sum_{k=1}^N \left[ {1\over n} \sum_{i=1}^n \{y(\mbs_i) - \hat{\text{prob}}(\mbs_i)\} ^2 \right]_k  \right)_m,
\end{align}
where, for given values of $m$ and $k$,
$$\widehat{\text{prob}}(\mbs_i) ={\exp\{\hat{A}(\mby(N_i))\} \over 1+\exp\{\hat{A}(\mby(N_i)) \}}.$$  For the GFLM,  
$$\hat{A}(\mby(N_i)) = \hat{\alpha} + \sum_{j=1}^{p} \hat{\beta}_j \e_j^{(i)},$$
while for our proposed SGFLM, 
$$\hat{A}(\mby(N_i)) = \log \left(\hat{\kappa}_i \over 1-\hat{\kappa}_{i}\right) + \hat{\eta} \sum_{j\in N_i} \{ y(\mbs_j) - \hat{\kappa}_j \},$$ with 
$$\log \left(\hat{\kappa}_i \over 1-\hat{\kappa}_i\right) = \hat{\alpha} + \sum_{j=1}^{p} \hat{\beta}_{j} \e_j^{(i)}.$$
To compare GFLM and SGFLM, we utilize $8,000$ datasets for GFLM without consideration of the location, whereas for SGFLM, we assume 20 independent realizations on a $20 \times 20$ regular lattice.

Estimation results for the $1,000$ Monte Carlo cases are shown in \autoref{table:repeating}.
Average estimates of $\eta$ were accurate for SGFLM, and had MSE values that were all less than $1\%$ of the true parameter value. 
As the GFLM lacks a spatial component, unlike SGFLM, we leave the results for the spatial dependence parameter $\eta$ blank.
Accuracy of MCLE for $\alpha$ with SGFLM was also quite good across all values of $\eta$, substantially better than that of GFLM using the quasi-likelihood approach of \cite{MS05}.  Those estimates had superior precision, with MSE ranging from about $10\%$ to $40\%$ of the corresponding values for GFLM.
Spatial structure appears to influence the estimation of $\beta$ for both models, with MISE increasing as the strength of spatial dependence increases, although this measure was consistently smaller for the model that accounts for such dependence (SGFLM) than the one that does not (GFLM). 
The integrated variance was perhaps more similar for the two models than the other performance measures.   
The overall fit criterion of FMSE was consistently smaller for SGFLM than for GFLM, and the improvement became greater as spatial dependence increased.  The FMSE for SGFLM was  $96\%$ of that for GFLM with the weakest dependence of $\eta=0.3$ but decreased to $94\%$, $86\%$ and finally only $74\%$ of the GFLM value as $\eta$ increased to $0.6$, $0.9$ and $1.2$, respectively. 

\begin{table}[b]
	\caption{Monte Carlo approximations of performance criteria for simulations on a $20 \times 20$ regular lattice with $\eta \in \{0, 0.3, 0.6, 0.9, 1.2\}$.}
	\footnotesize
	\centering
	\begin{tabular}{|c|ccccc|ccccc|}
		\hline
		& \multicolumn{5}{c}{GFLM} & \multicolumn{5}{|c|}{SGFLM} \\ \cline{1-11}
		$\eta$ & 0 &  $0.3$	&  $0.6$	&  $0.9$	&  $1.2$	& 0 & $0.3$	&  $0.6$	&  $0.9$	&  $1.2$	\\ \hline
		$\text{E}_M(\hat{\eta})$ & - & - & - & - & - & 	0.002& 0.300 & 0.601 & 0.903 & 1.199 \\ 
		$\text{MSE}_M(\hat{\eta})$ & - & - & - & - & - & 0.002 &0.002 & 0.002 & 0.002 & 0.001 \\ 
		$\text{E}_M(\hat{\alpha})$ & 0.227 &0.218 & 0.167 & 0.032 & -0.361 & 	0.000	&0.000 & -0.002 & -0.003 & -0.051 \\ 
		$\text{MSE}_M(\hat{\alpha})$  & 0.056 & 0.052 & 0.033 & 0.007 & 0.136 &	0.005	& 0.005 & 0.007 & 0.009 & 0.053 \\ 
		$\text{MISE}_M(\hat{\beta})$ &0.177	& 0.175 & 0.173 & 0.167 & 0.241 & 0.033&0.025 & 0.034 & 0.044 & 0.042 \\ 
		$\text{IV}_M(\hat{\beta})$ &0.027 & 0.027 & 0.026 & 0.026 & 0.099 & 0.033&0.025 & 0.034 & 0.043 & 0.042 \\ 
		$\text{FMSE}_M$ &	0.185 & 0.186 & 0.189 & 0.199 & 0.219 &	0.182 & 0.180 & 0.177 & 0.172 & 0.163 \\ 
		\hline
	\end{tabular}
	\label{table:repeating}
\end{table}	

Important focal points for inference with the SFGLM are the spatial dependence parameter $\eta$, and the parameter function $\beta$.
\autoref{table:inference} indicates that the coverage of confidence intervals for $\eta$ improved as the magnitude of this parameter increased, while the coverage of confidence bands for the parameter function $\beta$ remained consistently close to 0.96-0.97 across all cases.
Although our simulations included only $20$ copies of the data structure in each Monte Carlo case, the coverage results are quite satisfactory.

\begin{table}[t]
	\caption{Empirical coverages of confidence intervals for $\eta$ and confidence bands for $\beta$ based on MCLEs in simulations conducted on a $20 \times 20$ regular lattice with $\eta \in \{0, 0.3, 0.6, 0.9, 1.2\}$.}
	\centering
	\begin{tabular}{|c|ccccc|}
		\hline
		$\eta$ & 0 & $0.3$	&  $0.6$	&  $0.9$	&  $1.2$	\\ \hline
		$\text{CI}_M(\eta)$  & 	0.850	&0.837 & 0.843 & 0.869 & 0.913 \\ 
		$\text{CB}_M(\beta)$ & 	0.960 & 0.972 & 0.965 & 0.964 & 0.969 \\ 
		\hline
	\end{tabular}
	\label{table:inference}
\end{table}

\begin{figure}[b!]
	\centering
	\includegraphics[width=0.8\linewidth]{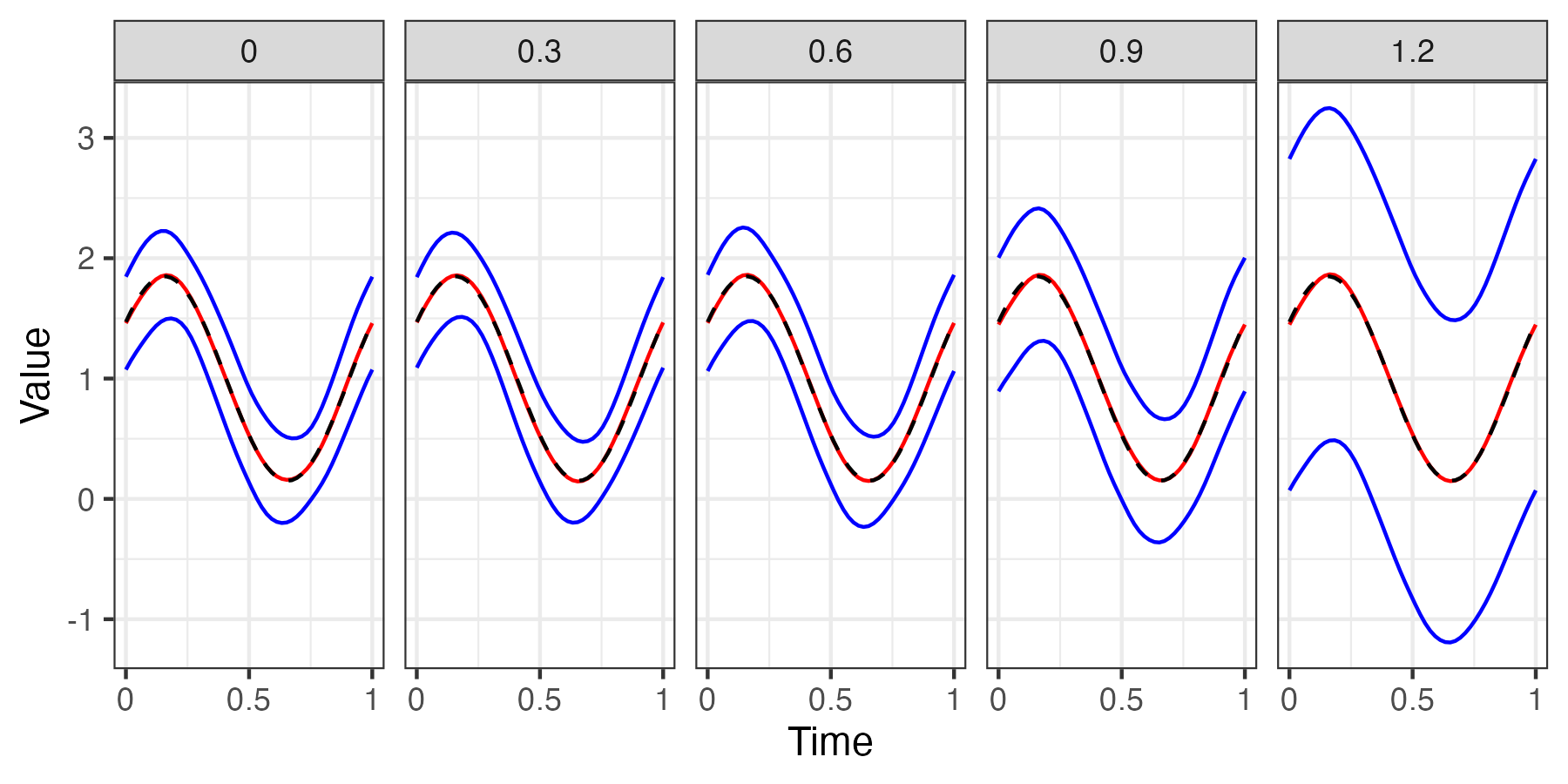}
	\caption{Confidence band for $\beta$ based on MCLEs from \autoref{conf_band} on $20 \times 20$ regular lattice with $\eta \in \{0, 0.3, 0.6, 0.9, 1.2\}$ in repeating lattice asymptotic context. The dotted black line is the true $\beta$, the red line is the average of $\hat{\beta}$, and the blue lines are the average of confidence bands of $\beta$.}
	\label{CB:repeating}
\end{figure}

\autoref{CB:repeating} presents the average estimate and average confidence band of $\beta$ over $1,000$ Monte Carlo cases. The confidence bands persistently include the true functions on average. Noticeably, the confidence band becomes substantially wider (again on average) when the spatial parameter $\eta$ becomes bigger. This trend can be attributed to the fact that the confidence bands described in \autoref{conf_band}, which involve ${G}_{22}^{(-1)}({\bm{\theta}})$, is connected to the entire parameter vector $\bm{\theta}$ including the spatial dependence parameter $\eta$.
Although we did not produce confidence bands for $\beta$ under the GFLM for every data set in the simulation, 
in 10 examined Monte Carlo cases, the bands were quite wide, resulting in the empirical coverage of exactly 1. 
When the spatial dependence parameter $\eta$ is at 0.3, average confidence bands ranged from about $-12$ to 14. 
In contrast, the average bands under the SGFLM, as shown in \autoref{CB:repeating}, ranged from about $-0.2$ to $2.2$. The average confidence bands under the GFLM also widen as the spatial parameter $\eta$ becomes bigger, which is the same trend as the average bands under the SGFLM in \autoref{CB:repeating}. 
Overall, our method exhibits strong performance with a relatively small number of repetitions ($N=20$), compared to the classical GFLM of \cite{MS05}, even under weak dependence ($\eta=0.3$), where the sample size for the latter is $n=8,000$, typically regarded as sufficiently large.

We used the first 20 functions $\{\phi_j\}_{j=1}^{20}$ from the trigonometric base when generating the functional covariate $X_i$ and fitting the model.
Additional simulation results using different basis functions for data generation and model fitting are provided in Section~S4 of the supplement.

In many problems, only a single observed field or lattice is available. In these settings, asymptotic inference relies on the context of an expanding lattice in which we assume that the lattice of spatial locations grows without bound, providing no replication but only a joint distribution that increases in dimension. 
We extend our approach to this expanding lattice framework, with further simulations presented in the supplement.
Briefly, we found that the selection of an appropriate truncation level for functional covariates may depend on the structure of the situation being considered, with a composite likelihood version of BIC  appearing superior to other possibilities for SGFLM applied to non-replicated lattice data.  Results were similar to those reported for the previous simulations with replicate lattices in that SGFLM outperformed other models when the data contained both spatial dependence and functional covariates.  Differences between models became more distinct as the degree of spatial dependence increased and as lattice size increased.

\section{An Application to Corn Yield and Maximum Temperature}\label{s6}
\begin{figure}[htbp]
	\centering
	
	\begin{minipage}{0.4\textwidth}
		\centering
		\includegraphics[width=\textwidth]{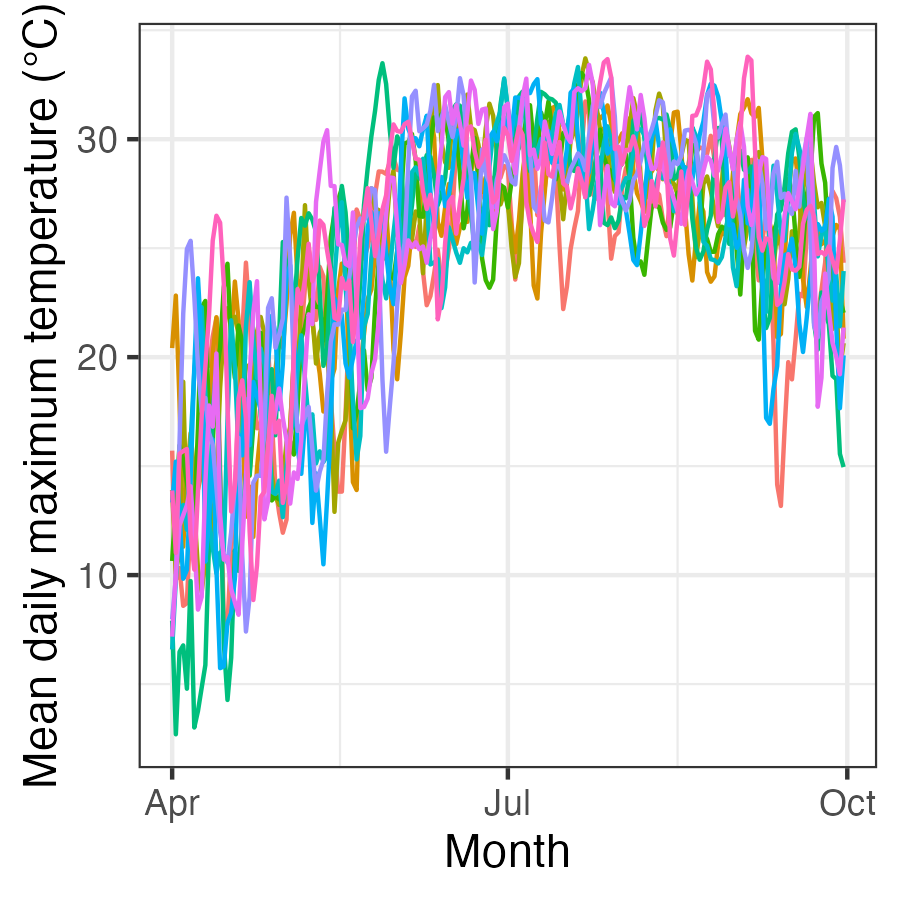}
	\end{minipage}
	\hspace{0.05cm}
	\begin{minipage}{0.58\textwidth}
		\centering
		\includegraphics[width=\textwidth]{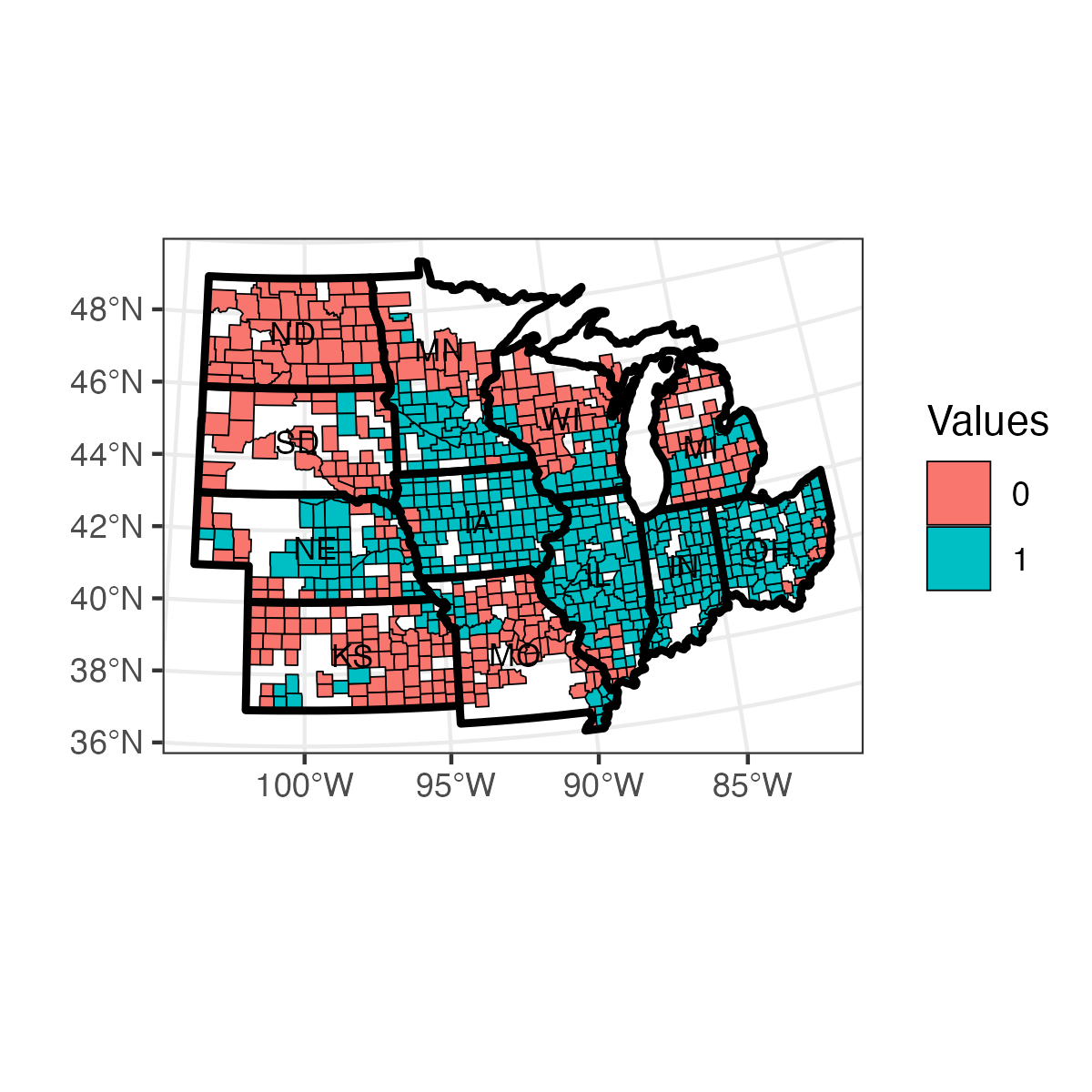}
	\end{minipage}
	\caption{In the left panel, each curve represents the average maximum temperature across all counties for each year. The right panel displays the binary responses for the year 2023 on a map.}
	\label{fig:data1_example}
\end{figure}

To illustrate the practical application of our model with data that contain multiple observed lattices, we consider data on annual corn yield and daily maximum temperature, spanning a time frame from 2014 to 2023.  For each year, functional covariates were defined as daily maximum temperature in Celsius, collected from April to September; these data were sourced from the National Centers for Environmental Information (NCEI).
Responses were considered to be annual corn yield data, measured in BU/ACRE, which were  obtained from the United States Department of Agriculture (USDA). Binary responses were created by taking the difference between yield for a county and the average yield across all counties.  If this difference was greater than zero, the binary response variable assumed a value of $1$, and was $0$ otherwise.  Using this binary response process served to mitigate much of the inter-annual variability in the magnitude of absolute corn yield, and the $10$ years of data were considered as repeated observations of the same lattice, providing the replication needed for repeating lattice asymptotics.
While possible temporal structure in response values is a legitimate concern, we will not explicitly incorporate any model components to address this issue here, assuming in our analysis that county-level corn yields are independent across years. 
Most of the factors that influence annual fluctuations in corn yields, such as the timing and level of rainfall and the amount of nitrogen mineralization by soil microbes, should not produce temporal patterns that could be reasonably represented by dependence in stochastic models.
\autoref{fig:data1_example} provides a descriptive example, displaying the average maximum temperature across all counties for each year alongside the binary responses for the year 2023 on a map. Notably, the average maximum temperature tends to peak in July to August, indicating a seasonal trend. To apply our model, we thus center the temperature observations as $X_i^{\mathrm{center}} = X_i - \bar{X}$.


The truncation level $p$ was determined as 6, based on the  AIC$_\text{c}$ criterion (\ref{AIC}).
The spatial dependence parameter was estimated as $\hat{\eta} = 0.67$ which,  accompanied by the 95\% confidence interval $(0.632, 0.709)$, indicates at least a moderate degree of spatial dependence. While a strict criterion to determine whether spatial dependence is large or small is not available, experience with this particular model suggests that degenerative behavior ensues if the spatial dependence exceeds 1.2. This suggests that a spatial dependence parameter approaching 1 implies a large spatial dependence, which leads to our conclusion that $\eta$ of about $0.7$ represents substantial dependence.

\begin{figure}[t]
\centering
\includegraphics[width=0.35\linewidth]{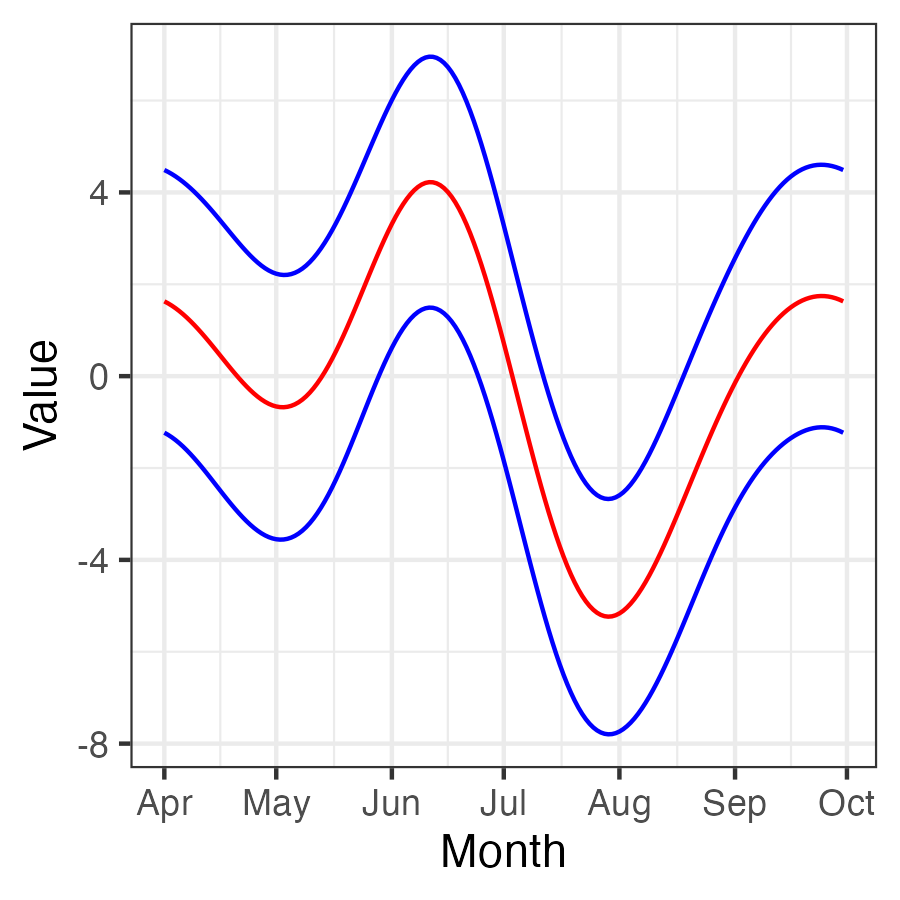}
\caption{The red line indicates the estimated parameter function $\hat{\beta}$ and the blue lines indicate the 95\% confidence band for $\beta$.}
\label{fig:data1_CB}
\end{figure}

\autoref{fig:data1_CB} illustrates the estimated parameter function $\hat{\beta}$ with 95\% confidence bands constructed by \autoref{conf_band}, while the intercept parameter $\alpha$ is estimated as $\hat{\alpha} = 10.911$.
The confidence band reveals two discernible relationships.
First, we found a positive relationship between maximum temperature and corn yield in June, likely due to the favorable temperature range for corn growth during this period. 
Conversely, from mid-July to mid-August, a negative relationship emerges between maximum temperature and corn yield. This relationship may be attributed to  high temperatures, reaching approximately 35 degrees Celsius (95 degrees Fahrenheit), that can occur during this period. Such extreme heat, if prolonged for more than a few days, reduces pollination, with a subsequent reduction in yield.

We also considered an application to a problem with a single observed spatial field in which the rate of new vaccinations against COVID is considered as a response, and the sequence of new infections over the previous weeks and months as a functional covariate.  Here we again detected spatial structure in responses as well as an impact of the functional covariate on those responses.  The rate of completed COVID vaccinations in counties of  the Midwest United States was positively related to the number of new infections over roughly the previous three to four weeks.  Although exact numerical results differed slightly, this relation was consistent over several time periods considered for definition of the functional covariate.  
Additional details and results are reported in Section~S5 of the supplement.

\section{Concluding Remarks}\label{s7}

We have developed a generalized linear model with a functional predictor and a scalar response, where spatial dependence originates from the response variable. 
Our work bridges the two important domains in Statistics, spatial statistics and functional regression, which represent a novel area of research.
Being able to account for such dependence improves the estimation of the regression parameter function, the usual focus of inference in regressions with functional covariates.  
The presence of the functional covariate process does not seem to degrade the estimation of  spatial structure, at least for binary response variables.  
We have provided solid theoretical results that allow inference under the asymptotic context of a repeating lattice, and those results appear to be applicable under a fairly modest number of repetitions of the data situation ($20$ in our simulations) as demonstrated in the numerical studies.  

Our methodology has been applied to a problem involving a response of annual corn yield and the sequence of daily maximum temperatures from April to September as a functional covariate.  This application involved repeated lattices with  a repeating size $N$ of 10.  Our analysis detected both spatial structure in corn yield and a positive relation between corn yield and maximum temperatures in June, suggesting that warm temperatures early in the growing season create favorable  conditions for corn growth. Conversely, for maximum temperatures from mid-July to mid-August, a negative relation with yield was evident, which can be attributed to extreme heat during this period adversely affecting corn pollination and yield.

A number of possible extensions of this work may be of interest. 
Models that include more complex spatial structure, such as directional
spatial dependence or models that incorporate both spatial and temporal dependencies, are natural topics to consider. 
Any of these modifications could be enacted by changing the definition of neighborhoods in the Markov random field approach.
It is also possible that one might encounter situations in which the functional covariate processes themselves exhibit spatial structure, either instead of or in addition to inherent spatial behavior in the response process. Because spatial structure in the covariate process should produce a certain amount of similar structure in responses, determining the relative contributions of covariate process dependence and direct dependence in responses should be a challenging problem, and this remains an area for future investigation. 
Another extension would be to use the functional regression model (\ref{model}) itself,  rather than relying on the $p$-truncated model. The $p$-truncated model assumes that the bias resulting from approximating the infinite-dimensional model is negligible. At least in the iid case, some studies have begun to examine this issue  \citep[cf.][]{DPZ12, ZG15}, and exploring a non-truncated functional model with spatial dependence remains an interesting avenue for additional investigation.

\begin{appendix}
%

\section{Assumptions}\label{s4.2}
We state the assumptions for our asymptotic theory.
Throughout this section, let $l_{c,[j]}^{'}({\bm{\theta}})$ and $l_{c,[j,j']}''({\bm{\theta}})$ denote the $j$-th and $(j,j')$-th elements of $l_{c}^{'}({\bm{\theta}}|\bm{y}_1)$ and $l_{c}''({\bm{\theta}}|\bm{y}_1)$, respectively. 
\leqnomode
\begin{align}\tag{A1}	\label{A1}
	&\limsup_N \max_{1\leq j \leq p_N+2} 
	\sup_{\tilde{\bm{\theta}}:\|\tilde{\bm{\theta}}-\bm{\theta}\|_2 \leq \Dt \sqrt{p_N \over N}{1\over b_N}  }
	\eo \left[|l_{c,[j]}^{'}(\tilde{\bm{\theta}})|^2\right] < \infty. 	
	\\&\tag{A2}	\label{A2}	
	\sup_{\tilde{\bm{\theta}}:\|\tilde{\bm{\theta}}-\bm{\theta}\|_2 \leq \Dt \sqrt{p_N \over N}{1\over b_N} }  
	N^{-1} \sum_{k=1}^N \|l_{c}''(\tilde{\bm{\theta}}|\bm{y}_k) - l_{c}''(\bm{\theta}|\bm{y}_k)\|_F
	\\& \hspace{5cm} =O_p \left(N^{-1/2} p_N b_N^{-1}\right).			\nonumber		
	\\&\tag{A3}	\label{A3}
	\limsup_N \max_{1\leq j,j' \leq p_N+2} 
	\eo \left[|l_{c,[j,j']}''({\bm{\theta}})|^2\right] < \infty.
\end{align}	

Assumptions~\eqref{A1}-\eqref{A3} are used to derive the existence and consistency of $\hat{\bm{\theta}}$ in \autoref{cons}.
Assumptions~\eqref{A1} and \eqref{A3} guarantee finite second moments of the first and second derivatives of the log composite likelihood, respectively, which ensures that $J(\bm{\theta})$ and $H(\bm{\theta})$ are well defined. These assumptions can be implied by a finite fourth moment of the regressor (cf.~Propositions~\ref{prop1}-\ref{prop2}), a common assumption in FDA. 
Assumption~\eqref{A2} embodies a degree of smoothness for the second derivative $l_c''$; the averaged discrepancy term in \eqref{A2} is bounded by $\sqrt{p_N}$ when $N^{-1/2}p_N^{1/2}b_N^{-1}$ goes to zero. This can be reduced to a boundedness condition on the regressor in the case of independence \citep[cf.][]{W11}.
\begin{align}\tag{A4}	\label{A4}
	&\tr\{G(\bm{\theta})\}=O(1).
	\\&\tag{A5}	\label{A5}
	\tr(G(\bm{\theta})^{-1})=O(N^{1/2} p_N^{-1}).
\end{align}

In the derivation of asymptotic normality, Assumptions~\eqref{A4}-\eqref{A5} are employed.
These assumptions, related to the Godambe information $G(\bm{\theta})$, are inspired by the properties of a covariance operator in an infinite dimensional space.
In the iid context, the Godambe information $G(\bm{\theta})$ becomes to the covariance matrix of the first derivative of the log likelihood, which is known as the Fisher information. As $p_N$ diverges to infinity, the Godambe information $G(\bm{\theta})$ is asymptotically close to a covariance operator, typically assumed to be bounded and trace class in FDA. 
Assumption~\eqref{A4} guarantees such boundedness of the trace of this covariance matrix.
Conversely, the inverse covariance operator, and thus its trace, is unbounded. Assumption~\eqref{A5} characterizes the growth rate of the trace to manage this unboundedness.
\begin{align}\tag{A6}	\label{A6}
	&\sum_{u,v,u',v'=1}^{p_N+2} \eo \left[
	l_{c,[u]}^{'}({\bm{\theta}}) l_{c,[v]}^{'}({\bm{\theta}}) l_{c,[u']}^{'}({\bm{\theta}}) l_{c,[v']}^{'}({\bm{\theta}})
	w_{uv}  w_{u'v'}
	\right]
	=o(N p_N^{-2} ).
	\\&\tag{A7}	\label{A7}
	\sum_{u_1,\dots, u_8}
	\eo\left[
	l_{c,[u_1]}^{'}({\bm{\theta}}) l_{c,[u_3]}^{'}({\bm{\theta}}) l_{c,[u_5]}^{'}({\bm{\theta}}) l_{c,[u_7]}^{'}({\bm{\theta}})
	\right]	
	\\& \quad \times
	\eo \left[
	l_{c,[u_2]}^{'}({\bm{\theta}}) l_{c,[u_4]}^{'}({\bm{\theta}}) l_{c,[u_6]}^{'}({\bm{\theta}}) l_{c,[u_8]}^{'}({\bm{\theta}})
	\right]
	w_{u_1u_2} w_{u_3u_4}   w_{u_5u_6}   w_{u_7u_8}  		\nonumber
	\\&\quad =o(N^2 p_N^2).							\nonumber
\end{align}	

Assumptions~\eqref{A6}-\eqref{A7} constitute mixed moment conditions which are the parallels to Conditions~(M3)-(M4) in \cite{MS05}.
These conditions are utilized in deriving the limiting distribution for the quadratic form of $\hat{\bm{\theta}}-\bm{\theta}$ in \autoref{normality_theta}. 
If we focus on the slope coefficient vector $\bm{\beta}=[\beta_0, \beta_1, \cdots, \beta_{p_N}]^\top$ and spatial dependence parameter $\eta$, then Assumptions~\eqref{A6}-\eqref{A7} can be substituted with other assumptions, such as Assumptions~\eqref{B1}-\eqref{B3}.

All the above conditions can usually be verified by moment assumptions and slow enough growth rate for $p_N$; two examples are given next.
Two examples in Propositions~\ref{prop1}-\ref{prop2} focus on logistic regression without or with spatial dependence.
Under some conditions, Assumptions~\eqref{A1}-\eqref{A3} in both examples are implied by the following assumptions \eqref{A1'}-\eqref{A3'}:
\begin{align}\tag{A1$^{\prime}$}	\label{A1'} 
	&\eo\|X_{1,1}\|_2^2 < \infty; 
	\\&\tag{A2$^{\prime}$}	\label{A2'} 
	\max_{1 \leq k \leq N} \max_{1 \leq i \leq n} \|X_{k,i} \|_2 = O_p(\sqrt{p_N}); \text{ and}
	\\&\tag{A3$^{\prime}$}	\label{A3'} 
	\eo\|X_{1,1}\|_2^4 < \infty.	
\end{align}
Assumptions~\eqref{A1'} and \eqref{A3'}, which are finite second and fourth moments of the regressor, respectively, are typical conditions in the FDA.
Assumption~\eqref{A2'} is implied by (A1) of \cite{W11}, which is a prevalent assumption in M-estimators with diverging dimensionality.
Propositions~\ref{prop1}-\ref{prop2} demonstrate that these can be verified by commonly assumed conditions.		

\begin{prop}\label{prop1}
	Suppose that we are interested in logistic regression with functional covariate without any spatial dependency. 
	It means that we have independent realizations $\{(\bm{X}_{k}, \bm{y}_k)\}_{k=1}^N$
	where $\bm{X}_k = X_{k,1}$ and $\bm{y}_k = y_{k,1}$ without having to consider the location.
	In this case, we will use $X_k$ and $y_k$ instead of $\bm{X}_k$ and $\bm{y}_k$ to denote the functional covariate and the scalar response, respectively. 
	The $p$-truncated functional logistic regression models can be represented as follows. For $k=1,\cdots, N$, 
	\begin{align*}
		y_k &\sim \mathsf{Bernoulli}(\text{prob}_k),
		\\ \log\left(\text{prob}_k \over 1-\text{prob}_k\right) &= \sum_{j=0}^{p_N} \beta_j \e_j^{(k)},			
	\end{align*}
	where $\e_j^{(k)} = \int X_k(t) \phi_j(t) dt$ and $\beta_j = \int \beta(t) \phi_j(t) dt$.
	In this case, the parameter vector of interest consists only of the slope coefficients vector without spatial dependence parameter $\eta$, i.e., $\bm{\theta} = \bm{\beta} = [\beta_0, \beta_1, \cdots, \beta_{p_N}]^\top$.
	
	We suppose that $H(\bm{\beta})$ has polynomial eigenvalue decay with $b_N = b(p_N) \asymp p_N^{-(1+\gamma)}$, $\gamma>0$.
	We further assume that 
	$|\int X(t) \beta(t) dt | < C$ almost surely, and $N^{-1/2} p_N^{3+\gamma} < 1$ (implying $N^{-1} p_N^{5+2\gamma} \rightarrow 0$ as $N \rightarrow \infty$).
	Then, Assumptions \eqref{A1}-\eqref{A7} are satisfied, where Assumptions \eqref{A1}-\eqref{A3} are implied by Assumptions \eqref{A1'}-\eqref{A3'}.
\end{prop}
%
%
\begin{prop}\label{prop2}
	Suppose that we are interested in logistic regression with functional covariate with spatial dependency as described in (\ref{model:ber}). 
	We suppose that $H(\bm{\theta})$ has polynomial eigenvalue decay with $b_N = b(p_N) \asymp p_N^{-(1+\gamma)}$, $\gamma>0$. 
	We further assume that $N^{-1/2} p_N^{5+2\gamma} < 1$, 
	and $N^{-1} p_N^4 c_N \rightarrow 0$ as $N \rightarrow \infty$.
	With Besag's original pseudo-likelihood and 4-nearest neighborhood, 
	Assumptions \eqref{A1}-\eqref{A7} are satisfied, where Assumptions \eqref{A1}-\eqref{A3} are implied by Assumptions \eqref{A1'}-\eqref{A3'}.
\end{prop}
The proofs of Propositions~\ref{prop1}-\ref{prop2} are provided in Section~S3 of the supplement.
In \autoref{prop1}, we considered an iid logistic regression with functional covariate, consistent with the setting in \cite{MS05}. Although the estimation methods differ---\cite{MS05}  used quasi-likelihood, while we focus on maximum likelihood estimation---we arrive at similar assumptions.
Even in more complex scenario of \autoref{prop2}, similar assumptions still hold.

We additionally establish the asymptotic normality of the quadratic form of $\hat{\bm{\beta}} - \bm{\beta}$
in \autoref{normality_beta}.
The following additional notation $Z(\bm{\theta})$ is introduced as
\begin{align*}
	Z(\bm{\theta}) = [h_{i2}^{(-1)} (G_{22}^{(-1)}(\bm{\theta}))^{-1} h_{2j}^{(-1)}]_{1 \leq i,j \leq 2} = [z_{uv}]_{1\leq u, v \leq p+2}
\end{align*}
to delineate the aforementioned conditions \eqref{B1}-\eqref{B3} that can replace Assumptions~\eqref{A6}-\eqref{A7}.

\begin{align}\tag{B1} 	\label{B1}
	&\sum_{u,v,u',v'=1}^{p_N+1}
	\eo\left[
	l_{c,[u]}^{'}({\bm{\theta}}) l_{c,[v]}^{'}({\bm{\theta}}) l_{c,[u']}^{'}({\bm{\theta}}) l_{c,[v']}^{'}({\bm{\theta}})
	z_{uv}  z_{u'v'}
	\right]
	=o(N  p_N^{-2}).
	\\&\tag{B2} 	\label{B2}
	\sum_{u,v,u',v'=1}^{p_N+1} \sum_{s,t,s',t'=1}^{p_N+1} 
	\eo\left[
	l_{c,[u]}^{'}({\bm{\theta}}) l_{c,[u']}^{'}({\bm{\theta}}) l_{c,[s]}^{'}({\bm{\theta}}) l_{c,[s']}^{'}({\bm{\theta}})
	\right]
	\\& \hspace{5cm} \times z_{uv} j_{vv'} z_{v'u'}  z_{st} j_{tt'} z_{t's'}  \nonumber
	=o(N  p_N^{-2}).														\nonumber
	\\&\tag{B3} 	\label{B3}
	\sum_{all \, u,v}			
	\eo\left[
	l_{c,[u_1]}^{'}({\bm{\theta}}) l_{c,[u_2]}^{'}({\bm{\theta}}) l_{c,[u_3]}^{'}({\bm{\theta}}) l_{c,[u_4]}^{'}({\bm{\theta}})
	\right]
	\\& \quad  \times
	\eo\left[
	l_{c,[v_1]}^{'}({\bm{\theta}}) l_{c,[v_2]}^{'}({\bm{\theta}}) l_{c,[v_3]}^{'}({\bm{\theta}}) l_{c,[v_4]}^{'}({\bm{\theta}})
	\right]
	z_{u_1v_1} z_{u_2v_2}  z_{u_3v_3} z_{u_4v_4} 							\nonumber
	\\&\quad =o(N^2 p_N^2).															\nonumber
\end{align}
Assumptions~\eqref{B1}-\eqref{B2} can be a replacement of Assumption~\eqref{A6} to handle the non-identity nature of $Z(\bm{\theta}) J(\bm{\theta})$, unlike the case of $W(\bm{\theta}) J(\bm{\theta}) = I_{p_N+2}$.
We can view Assumption~\eqref{B3} as a version of Assumption~\eqref{A7} for the case of excluding the spatial dependence parameter $\eta$ from the entire parameter vector $\bm{\theta}$. 

In the case of inference of $\eta$, we can impose a different assumption for the asymptotic normality of $\hat{\eta}$ in \autoref{normality_eta}.
\begin{align}\tag{E1}	\label{E1}
	\eo[(e_1^\top H(\bm{\theta})^{-1} l_c'(\bm{\theta}|\bm{y}_k))^4] = o\left(N \left( G_{11}^{(-1)}(\bm{\theta}) \right)^2\right).
\end{align}
Here, $e_1$ represents the $(p_N+2)$-dimensional unit vector for which the first element is $1$ and the other elements are all $0$.
In \autoref{normality_eta}, we demonstrate the asymptotic normality of $\hat{\eta}$ under Assumption~\eqref{E1} instead of the original conditions \eqref{A6}-\eqref{A7}. 
Unlike Theorems~\ref{normality_theta}-\ref{normality_beta}, we do not necessarily rely on the quadratic form for the asymptotic inference of $\eta$.
In particular, under Assumption~\eqref{A5}, the fourth mixed moment in Assumption~\eqref{E1} is bounded by $o(N^2p_N^{-2})$, which is a weaker bound than the upper bound in Assumption~\eqref{A6}. 

Lastly, we establish that $G(\bm{\theta})$ can be replaced by $\hat{G}(\hat{\bm{\theta}})$ for inference in Theorems~\ref{normality_theta}-\ref{normality_eta}.  This substitution requires additional assumptions.
\begin{align}\tag{G1}	\label{G1}
	&\sup_{\tilde{\bm{\theta}}:\|\tilde{\bm{\theta}}-\bm{\theta}\|\leq \Dt \sqrt{p_N \over N}{1\over b_N} }  
	N^{-1} \sum_{k=1}^N \|l_{c}'(\tilde{\bm{\theta}}|\bm{y}_k)l_{c}'(\tilde{\bm{\theta}}|\bm{y}_k)^\top - l_{c}'(\bm{\theta}|\bm{y}_k)l_{c}'(\bm{\theta}|\bm{y}_k)^\top \|_F
	\\& \hspace{7cm} =O_p(N^{-1/2} p_N b_N^{-1}).		\nonumber
\end{align}		
\begin{align}\tag{G2}	\label{G2}
	\limsup_N G_{11}(\bm{\theta})^{-1} < \infty.
\end{align}
Assumption~\eqref{G1} characterizes a degree of smoothness for $l_{c}' l_{c}'^\top$, which is similar to Assumption~\eqref{A2}, while Assumption~\eqref{G2} serves as a technical condition for \autoref{normality_beta} to deal with the block matrix inversion, $(G_{22}^{(-1)}(\bm{\theta}))^{-1}$.

\end{appendix}
\begin{supplement}
\stitle{Supplement to ``Generalized linear models with spatial
	dependence and a functional covariate"}
\sdescription{This supplement includes the proofs of theorems and lemmas for the main results, as well as the proofs of the propositions detailed in \autoref{s4}. It also presents additional simulation and real data analysis results under the expanding lattice context. (.pdf file).}
\end{supplement}


\bibliographystyle{imsart-nameyear} 
\bibliography{reference_paper}       

%
%
%
%

\end{document}